\documentclass[11pt]{article}
\usepackage{graphicx}
\usepackage{amsfonts}
\usepackage{amsmath}
\usepackage{amssymb}
\usepackage[margin=2.54cm]{geometry}

\begin{document}

\title{Solvability by semigroup : Application to seismic imaging with complex decomposition of wave equations and migration operators with idempotents}

\author{August Lau and Chuan Yin \\
\\ Apache Corporation \\
        2000 Post Oak Blvd., Houston, Texas 77056 \\
        \\
        Email contact: \texttt{chuan.yin@apachecorp.com}}

\date{January 28, 2011}

\maketitle

\begin{abstract}

The classical approach of solvability using group theory is well known and one original motivation is to solve polynomials by radicals.   Radicals are square, cube, square root, cube root etc of the original coefficients for the polynomial.   A polynomial is solvable by radicals if the permutation group is solvable.  This is exact solvability via group theory.  With modern computers, we might need to relax our definition of exact solvability and move towards practical solvability. We will address seismic imaging as an example of practical solvability by semigroup theory.  The difference between semigroup and group is that the semigroup operators do not have to be invertible as in group operators.   Using the metaphor of complex decomposition,  we will decompose an operator into simple part and complex part.   The simple part of the operator is solvable by numerical methods.  The complex part of the operator is interpretable but not numerically solvable.  It is sometimes called the evanescent energy in geophysics.
\end{abstract}

\section*{Introduction}

Many numerical problems in differential equations and probability can be cast as operators acting on Hilbert space.  In this paper,  we can treat this more abstract operator theory by using $n \times n$ matrices as a surrogate for the more general operator acting on Hilbert space.  Instead of factoring polynomials in exact solvability, we can use the same concept of group theory to talk about solving the matrix by diagonalization of the matrix with an orthogonal group of matrices.  This can be viewed as solvability by group theory similar to solving polynomials with permutation groups.   If we use Gaussian elimination or similar techniques to diagonalize a matrix,  each operation on the original matrix could be thought of as an invertible transformation of the original matrix.  For large seismic surveys which are in many terabytes of input data,  it is untenable to use the full matrix and diagonalize the matrix.  One way to generalize the idea is to use semigroup theory where the operators are not invertible like rotations in group theory.   We can think of taking an operator $A$ and write it as sum of idempotent operators, i.e., $A$ is a formal sum of $P1+P2+P3+\dots $,   where $A$ is the operator and $P_{j}$ is an idempotent with $P_{j}P_{j} = P_{j}$.  Idempotents are operators $P$ such that $PP = P$.  The more familiar notion of idempotent is eigenfunctions or projections or diagonal blocks or fixed points or spectral method in numerical analysis.  Decomposition into idempotents is a way to achieve stability.  We will use wave equation in seismic imaging to demonstrate solvability by semigroup theory.

\section*{Seismic imaging}

Migration of seismic data can be classified as operators with predetermined basis and without predetermined basis.   Migration is the geophysical terminology to map the input recorded data into a seismic imaged cube using wave equation.

The first category includes $fk$ migration and phase shift migration which uses Fourier basis.  Its advantages are speed and no limitation of bandwidth or dips.  The disadvantages are inability to handle large velocity variations and inaccuracy of interpolation over a large seismic survey.  

The second category includes finite difference method which includes downward continuation and reverse time migration (RTM) and integral method (Kirchhoff or Green's function).  The advantage is the ability to handle large velocity variations but could have frequency and dip limitations due to the practicality of using finite difference over a large survey.   Even though technically finite difference or Kirchhoff method does not use a predetermined basis,  it requires a predetermined order of approximation for the whole survey.

To this end,  we will employ an idempotent method which  honors large velocity variation and all frequencies and dips.  It can also accommodate both 1-way and 2-way wave propagation.  In this paper,  we will address the theory and practice of idempotent method.  We will not discuss the numerical implementation of the method. 

\section*{Methodogy}

We will first introduce some mathematical concepts in semigroup and group theory.  In order to make it less abstract,  we will think of operators as matrices since matrices are a good surrogate to discuss abstract ideas.   If we take all $n \times n$ matrices under matrix multiplication as a whole system,  it is a semigroup and it has the semigroup property of associative law $A(BC) = (AB)C$.  This system includes both invertible and noninvertible matrices.   We will come back to the idea of semigroup and idempotents in semigroups.

If we restrict a subset to all invertible matrices,  then it is a group since each $n \times n$ invertible matrix  $A$ by definition will have a $B$ so that $AB = BA = I$ the identity matrix.  Two matrices are equivalent if $A = UBV$ where $U$ and $V$ are invertible.   A desirable property is to write a matrix as  $A = UDV$ where $D$ is a diagonal matrix.  More generally,  it is desirable to have $A = UPV$ where $P$ is an idempotent with $PP = P$.  

In a group,  the only idempotent is the identity.  In other words,  when we square or cube a matrix,  it gives a different matrix compared to $A$.  It is deemed desirable to iterate to get a new matrix which is different.  However,  it is in general difficult to know if the iterated matrices converge to a meaningful answer.  It could converge numerically but it does not mean that it converges to a geometrically/geologically meaningful solution.

An idempotent in a semigroup has been viewed as useless since $AA=A$.  It means that nothing new happens.  But it is the most stable matrix since nothing new could occur.  This is the point of using idempotent which is similar to projection operators in numerical implementations.  An idempotent projects the whole vector space of $n$-dimensional Euclidean space into a subspace.

The intuitive notion of an idempotent in migration operator is to select the appropriate projection to honor both the wave equation and the complicated velocity model.  This is like "lumping" wave equation and velocity model into ONE operator.   The idempotent maintains stability and adapts to the large velocity variation (Figure 1).

\section*{A comment on physics versus mathematics}

A typical seismic imaging experiment in oil exploration is to have compressional sources (dynamite, airguns, vibrating sources etc) at the surface of the earth and geophones/hydrophones to measure the energy coming back from the reflections of the earth layers.  Our goal is to create a seismic cube which can be interpreted as geologic layers to find oil and gas.  One such method is downward continuation of the wavefield.

We will use a wave equation without any loss of energy.  So for the physics,  we will assume total conservation of energy.   In order to image the earth layers,  we need to downward continue the wavefield recorded at the surface to deeper layers where the reservoir containing oil and gas resides.   Downward continuation is the numerical method to extrapolate the wavefield recorded at the surface to the deeper layers in the earth.  An acoustic or elastic wave equation is used to define the extrapolator into the earth.   

Even though the physics is not lossy,  the mathematics is lossy.   It requires that evanescent energy has to be removed at every depth step as we extrapolate the wavefield deeper into the earth.  Evanescent energy is the numerically unstable part of the recording which grows exponentially in each depth step.   So even if the physics is conservative,  the mathematics has to eliminate part of the recorded energy and hence the extrapolation is lossy.  It is not a reversible process.  If we extrapolate into the earth and then undo the extrapolation back to the surface,   the wavefield without the evanescent energy is not the same as the recorded wavefield.  

The simple part of the wavefield is preserved but the complex part of the wavefield is eliminated by the extrapolation method.

\section*{Complex decomposition of operators}

We have used the metaphor of complex decompostion of data into simple part and complex part  (see Lau et al SEG 2008).  The simple part of the data is numerically explainable and the complex part of the data is the left over part which is interpretable but is not explainable by numerical methods.   This idea was used for data decomposition.  With operator decomposition,  the simple part is the part of the operator without evanescent energy.   The complex part is the evanescent energy.

\subsection*{Decomposition of operators (with predetermined basis)}

The simplest way to decompose an operator is to use a predetermined basis like Fourier basis.  The advantage of predetermined basis is to convert a differential equation into an algebraic equation with orthogonal basis like $fk$ ($f$ is frequency and $k$ is wavenumber).  The evanescent wave is just filtering certain dips (i.e. certain $fk$ numbers)(Figure 2).  The disadvantage is the difficulty of handling large velocity variations.

Any predetermined basis forces an operator decomposition into predetermined geometry like Fourier basis.   The predetermined basis maps the operators into fixed shapes given by the basis elements.

\subsection*{Decomposition of operators (without predetermined basis)}

The second method is to lump the derivatives with the velocity field so it is more sensitive to the velocity variations.  Finite difference and Kirchhoff extrapolation are such methods but they can have practical limitations for frequencies and dips.   Integral methods like Kirchhoff extrapolation (see Berryhill) are more geometrically motivated.  An excellent numerical treatment can be found in Sandberg and Beylkin 2009 where positive eigenvalues are viewed as evanescent energy.  The negative and zero eigenvalues are kept as propagating waves in the extrapolation operator.  This is another way to separate the wavefield into simple part and complex part.

We use wave field extrapolation of 1D acoustic wave equation to demonstrate decomposition of operators without pre-determined basis. 1D acoustic wave equation can be written as follows
\begin{equation}
\ddot{p}=\rho c^2 \frac{\partial}{\partial z} \left( \frac{1}{\rho}\frac{\partial p}{\partial z} \right),
\end{equation}
where $p(z,t)$ represents pressure, $\rho(z)$ the bulk density, and $c(z)$ the velocity. A dot above a variable denotes differentiation with respect to time. Fourier transform Eq.(1) with respect to time, and write in the form of two coupled first-order equations (see e.g., Richards 1971)
\begin{equation}
\dfrac{d}{dz} \begin{bmatrix} \hat p \\ \\ \dfrac{1}{\rho} \dfrac{d \hat p}{dz} \end{bmatrix}
= 
\begin{bmatrix} 
0 & & \rho \\  & \\
-\dfrac{\omega^2}{\rho c^2} & & 0 \\
\end{bmatrix}
\begin{bmatrix} \hat p \\ \\ \dfrac{1}{\rho} \dfrac{d \hat p}{dz} 
\end{bmatrix},
\end{equation}
where $\hat p(z,\omega)$ is the Fourier transform of $p(z,t)$. Equation (2) is written consistently with the continuity conditions of continuum mechanics which require that both the pressure and displacement remain continuous across all possible interfaces in the medium. 

We can re-write Eq.(2) into matrix form,
\begin{equation}
\dfrac{d}{dz} \mathbf{f} = \mathbb{A} \mathbf{f},
\end{equation}
which has the following solution when $\mathbb{A}$ is independent of $z$,
\begin{equation}
\mathbf{f} =  e^{z \mathbb{A}} \mathbf{f}_{0},
\end{equation}
where $\mathbf{f}_{0} = \mathbf{f}(z=0)$.

A traditional way of analysing Eq.(4) is by Taylor series expansion 
\begin{equation}
\mathbf{f}(z) = \left( \sum_{k=0}^{\infty} \dfrac{1}{k!} (z \mathbb{A})^{k} \right) \mathbf{f}_{0} 
=(\mathbb{I}+z \mathbb{A} + \dfrac{1}{2}z^2 \mathbb{A}^{2}+ ... .) \mathbf{f}_{0} 
\end{equation}
Assuming convergence, one can then truncate the series and apply numerical analysis tools to solve the equation.

Another way of further analyzing Eq.(4) is to examine eigenvalues and eigenvectors of $\mathbb{A}$, i.e., $\mathbb{A} = \mathbb{VDV}^{-1}$, where $\mathbb{V}$ is the eigenvector matrix, $\mathbb{D}$ diagonal matrix of the eigenvalues.

Eq.(4) becomes 
\begin{equation}
\mathbf{f} =  \mathbb{V} e^{z \mathbb{D}} \mathbb{V}^{-1} \mathbf{f}_{0},
\end{equation}
where the diagonal eigenvalue matrix is
\begin{equation}
\mathbb{D}=
\begin{bmatrix}
-i \dfrac{\omega}{c} & & 0 \\ \\ 0 & & i \dfrac{\omega}{c} \\
\end{bmatrix}, 
\end{equation}
and the eigenvector matrix and its inverse are respectively, 
\begin{equation}
\mathbb{V}=
\begin{bmatrix}
i \dfrac{\rho c}{\omega} & & -i \dfrac{\rho c}{\omega}  \\ \\ 1 & & 1 \\
\end{bmatrix}, 
\end{equation}
and 
\begin{equation}
\mathbb{V}^{-1}= \dfrac{1}{2}
\begin{bmatrix}
-i \dfrac{\omega}{\rho c} & & 1  \\ \\ i \dfrac{\omega}{\rho c}  & & 1 \\
\end{bmatrix}. 
\end{equation}
We can then define a projection of $\mathbf{f}$ into $\mathbf{g}$ by
\begin{equation}
\mathbf{g} = \mathbb{V}^{-1} \mathbf{f}.
\end{equation}
Notice the projection depends on the velocity $c$ and the density $\rho$.

Eq.(6) then becomes
\begin{equation}
\mathbf{g} =  e^{z \mathbb{D}} \mathbf{g}_{0}.
\end{equation}
The operator,
$$\mathbb{L} = e^{z \mathbb{D}},$$ 
in Eq.(11) can be written in more general form as
\begin{equation}
\mathbb{L} =  \mathbb{P}_{1} + \mathbb{P}_{2},
\end{equation}
where
\begin{equation}
\mathbb{P}_{1} =  \mathbf{e}_{1} e^{z \mathbb{D}},
\end{equation}
and
\begin{equation}
\mathbb{P}_{2} =  \mathbf{e}_{2} e^{z \mathbb{D}},
\end{equation}
in which
\begin{equation}
\mathbf{e}_{1}=
 \begin{bmatrix}
1 & & 0 \\ \\ 0 & & 0 \\
\end{bmatrix}, 
\end{equation}
and 
\begin{equation}
\mathbf{e}_{2}=
 \begin{bmatrix}
0 & & 0 \\ \\ 0 & & 1 \\
\end{bmatrix}, 
\end{equation}
are idempotents, which corresponds to upward-propagrating and downward-propagating components of the wavefield, respectively. This simple approach of decomposing 1D wave equation operator can potentially help in reducing numerical instabilities in wave field downward continuation and seismic imaging.

\section*{Topological complexity}

There are geologic layers with fractures and faults which are beyond wave equation or any differential equation or probability.   This is a more complicated topic since no equation could be written when the topology like fractures is too complex.  We might need to consider qualitative mathematics like computational topology which measures homology or cohomology of the geology.  One such approach is to compute Betti numbers of the homology or cohomology group.  See Kaczynski et al,  Lau and Yin in arXiv 2010 on "L0+L1+L2 optimization".

\section*{Conclusion} 

Solvability by semigroup is a way to use noninvertible operators to approximate the original operator.   We have chosen in this paper to use idempotents $P$ where $PP = P$.   Semigroup theory is not restricted to idempotents but idempotents are easier to demonstrate since they resemble diagonal matrices (see Davies, Wilansky).  They are also familiar to physicists in terms of density matrix with 0 or 1 eigenvalues (see Bowler et al) viewed as basic building blocks.  

Multiplication of noninvertible operators becomes "less energetic" or contractive or lossy or more limited.   Addition of noninvertible operators becomes "more energetic" or expansive or generating larger operator.  Toggling between multiplication and addition of the idempotents gives us the stable part (multiplicative) and the generative part (addition).  This is the essence of downward continuation in wavefield extrapolation.  The schematic is captured with the diagram in the end (Figure 1). 

The simple part of the wavefield extrapolation operator is written as the sum of idempotent operators  ( $A = P1+P2+P3+\dots $ ) where $A$ is the operator and $P_{j}$ is an idempotent with $P_{j}P_{j} = P_{j}$ and $P_{j}P_{k} =0$ if $P_{j}$ and $P_{k}$ are different.  We dropped the phrase "equivalent to idempotent" to emphasize the importance of idempotents.  The simple part of the operator is the propagating operator to extrapolate the wavefield.  It can be studied through semigroup of operators. The complex part of the operator is the evanescent energy which should not be propagated.  The complex part needs qualitative mathematics like topology to classify and understand.

\bibliographystyle{plain}
\bibliography{solvability_semigroup}

\begin{figure}
\centering
  \includegraphics[width=6.5in]{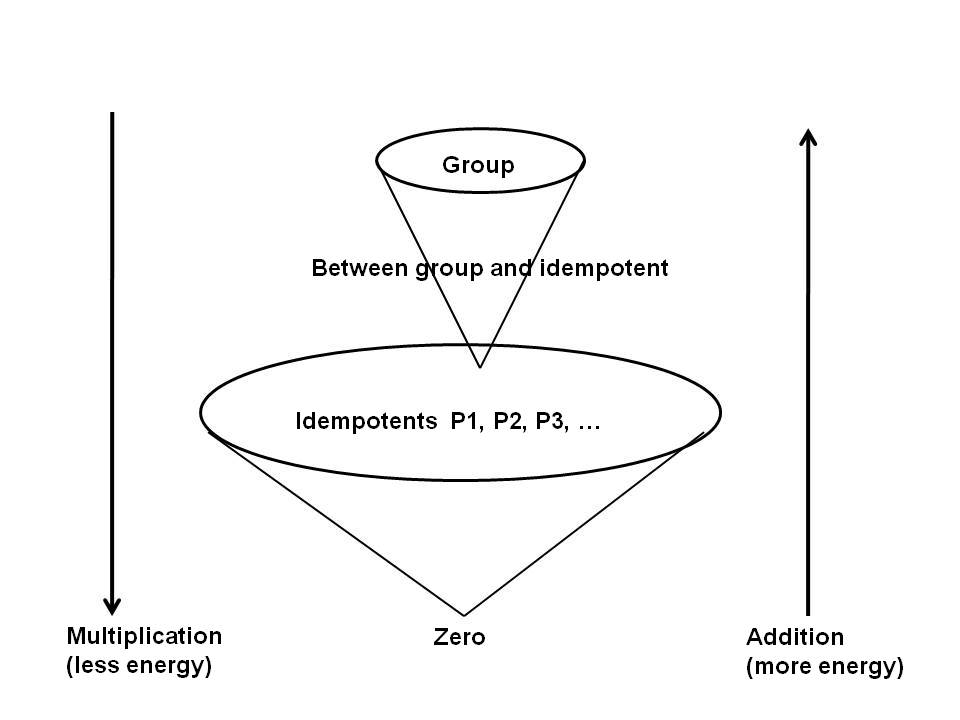}
\caption{A lower energy, stable state can be reached by multiplication of idempotents.}
\label{fig1}
\end{figure}

\begin{figure}
\centering
  \includegraphics[width=6.5in]{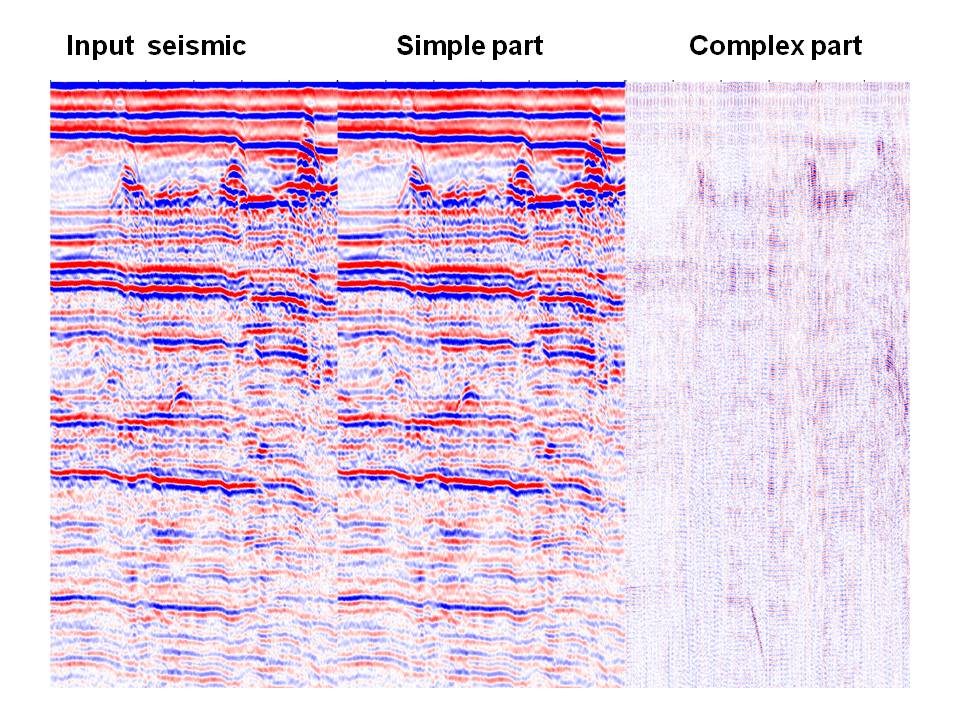}
\caption{An example of decomposing a 2D seismic section into simple and complex parts, using Fourier basis.}
\label{fig2}
\end{figure}

\end{document}